# A Dynamic Model for Sharing Reputation of Sellers among Buyers for Enhancing Trust in Agent Mediated e-market

Vibha Gaur[1], Neeraj Kumar Sharma[2], Punam Bedi[3]

[1] Department of Computer Science, University of Delhi
Delhi, India
3.vibha@gmail.com

[2] Department of Computer Science, University of Delhi
Delhi, India
neerajraj100@gmail.com

[3] Department of Computer Science, University of Delhi
Delhi, India
pbedi@du.ac.in

**Abstract**
Reputation systems aim to reduce the risk of loss due to untrustworthy participants. This loss is aggravated by dishonest advisors trying to pollute the e-market environment for their self-interest. A major task of a reputation system is to promote and encourage advisors who repeatedly respond with fair advice and to apply an opinion filtering or honesty checking mechanism to detect and resist dishonest advisors. This paper provides a dynamic approach to compute the aggregated shared reputation component by filtering out unfair advice and then generating the aggregated shared reputation value. The proposed approach is dynamic in nature as it is sensitive to the behaviour of advisors, value of the current transaction and encourages the cooperation among buyers as advisors. It provides incentive to honest advisors in lieu of repeated sharing of honest opinion by increasing the weight of their opinion and by making the increase in the reputation of honest advisors monotonically proportional to the value of a transaction.

## 1. Introduction

The burst in Internet connectivity around the globe has resulted in the enormous increase in the popularity of e-commerce for online buying and selling. This next generation technique of online business and consumer behaviour has associated risks as it relies on cooperative user behaviour [7]. Ensuring cooperative behaviour among participants is a challenging task undertaken by online reputation systems given the facts that participating users are autonomous and self-interested, behaving only in their best interests. In the design of a robust reputation system, different reputation management processes like collection and aggregation of reputation should require safeguarding against a variety of threats like biased/unfair opinions due to collusion between the advisor, other buyers and target seller agent.

Buyer agents that provide opinions about a target seller agent play a vital role in the success of a reputation system because based on their opinion the value of shared reputation component will be computed. A false opinion may result in future transactions with untrustworthy seller agent as the source buyer agent that has requested for advice might be misinformed and may form a wrong assessment of the target agents' reputation [5]. Hence opinions with hidden motives and malicious intentions can harm reputation systems by artificially enhancing the reputation of malicious participants or by artificially lowering the reputation of honest participants resulting into possible failure of the system. Therefore, mechanisms to filter out dishonest advice are fundamental part of any reputation sharing strategy, and are an integral part of the success of any online reputation system.

One of the major objectives of a reputation system is to promote and encourage advisors who provide fair advice for enhancing trustworthiness among participants in the e-market environment. An honest opinion helps in narrowing the gap between the shared reputation score and the actual trustworthiness of the target agent under consideration. Reputation systems aggregate opinions from other community members playing the same role, say buyers role, in a meaningful way to compute the peer's reputation. The trade-off in the opinion gathering process is between the efficiency of using available reputation information and vulnerability to false ratings and deceitful behaviour. A robust reputation system should be resistant to the participants' efforts who try to maliciously tamper with reputation information by presenting false information to reputation seekers. Various rating misbehaviour that are observed during the opinion gathering process are listed below.

- Individual Unfair Ratings: A particular advisor provides unfairly high or low ratings without collaborating with other agents. This type of rating may result from advisors' personality, carelessness or randomness in rating behaviour [3, 20].

- Collaborative Unfair Ratings: A group of agents collude to provide unfairly high or low ratings to boost or downgrade the overall rating of the target agent [1, 2, 10, 12, 19, 20]. Two subtypes of collaborative unfair ratings are known as Ballot Stuffing (BS) and Bad Mouthing (BM). In BS, a number of buyers give inflated reputation ratings of the target seller agent with the purpose to collectively boost the reputation of the target agent at the raters' end by effectively launching a good reputation attack. In BM, a number of buyers provide unfairly low reputation ratings of the target seller agent with the purpose to conspire against the target seller agent, thereby hurting its reputation.

Compared with collaborative unfair ratings, individual unfair ratings usually cause much less damage [4]. First, individual high ratings and individual low ratings can cancel out each other. Second, the number of individual unfair ratings should be statistically much less than the number of normal ratings. Therefore, this paper concentrates on filtering out collaborative unfair ratings.

This paper utilises Reinforcement learning (RL) for modeling the reputation of advisors. RL is a machine learning technique that deals with what an agent should do in every state that it can be and how to map situations to action, in order to maximize the long term reward. The learner is not told which action to take, but instead must discover which actions yield the most reward by trying them. In some cases, actions may affect not only the immediate reward, but also all subsequent rewards. These two characteristics, i.e. trial-and-error search and delayed reward are most important distinguishing features of RL.

The proposed model for sharing reputation of seller agents amongst buyer agents is based on the dynamic reputation model in e-market [15, 16, 17, 18] in which the reputation of a seller agent is computed using reinforcement learning. Reputation of a participant in this model is composed of two components: Individual Reputation (IR) component and Shared Reputation (SR) component. Individual Reputation (IR) is based on the direct personal experience of the buyer with the seller, whereas Shared Reputation (SR) represents the advice/opinion of other buyer agents who have some experience of having previous transactions with the seller agent. This paper provides a strategy to share reputation of a seller agent (henceforth referred to as target seller agent) among buyer agents. It also deals with the problem of providing unfair reputation rating by a set of buyers in the role of advisors to the source buyer agent who has solicited opinions about a seller agents' reputation from other buyer agents in the e-market.

The proposed methodology in this paper contains two key elements: filtering of unfair ratings and establishing/updating advisors' reputation. Filtering of unfair ratings involves getting rid of those opinions that do not appear to be trustworthy by virtue of being either artificially inflated or artificially reduced. The second key element is reputation establishment of the advisors, as an advisors' reputation that is based on its history of previous interactions reveal information about its past behaviour as an honest/dishonest advisor. Further, an expectation that the nature of its current advice will be visible in the future may deter moral hazard in the present, that hazard being the temptation to cheat or exert low effort [4] and would create an incentive to honestly perform up to its ability. Knowledge about the trustworthiness of advisors would reduce the influence of the untrustworthy advisors on the system by avoiding the advice of dis-reputed advisors. Further, as the unfair advice is filtered out, different reputation ratings of the target seller agent received as opinion from other buyer agents must be aggregated to form a single aggregated shared reputation component value. The advice from other buyers is filtered by using the concept of moment about an arbitrary point [8, 14] that compares the set of received opinions with the reputation value representing the actual individual experience of the source buyer agent. The proposed scheme provides an incentive to honest advisors by allocating reputation to advisors for their honest opinions in the reputation sharing process and also by making weight of an advisors' opinion monotonically proportional to number and percentage of honest past opinions as a reward for successive fair opinions in an environment where honesty is at premium.

The rest of this paper is organized as follows. Various methods of sharing reputation in reputation models from the literature are presented in section 2. Section 3 presents a trust enhancing dynamic reputation sharing methodology among buyer agents. Section 4 comprises of a case study. The paper concludes with Section 5.

## 2. Related Work

Reputation models are an important component of e-market, help building trust and elicit cooperation among loosely connected and geographically dispersed economic agents [13]. A number of research works from literature have recognized the importance of reputation management in online systems and a number of solutions have been proposed to promote cooperation among agents for sharing of reputation information for a more accurate reputation computation. These models employ various strategies to filter out unfair/dishonest opinions as discussed ahead in this section.

Beta Reputation System (BRS) [1] estimates reputation of seller agents by using a probabilistic approach using beta probability density functions. It combines the ratings of a seller agent being provided by multiple advisors by accumulating number of good ratings and number of bad ratings. To handle unfair opinions, BRS filter out ratings that are not in the majority among others by using Iterated Filtering approach. However, this approach is effective when majority of ratings are fair and it does not combine buyers' personal experience with the advisors' feedback.

TRAVOS [19] is a trust and reputation model that models an agents' trust in a partner by taking into account past interactions. It deals with inaccurate reputation advice by first estimating the accuracy of the current reputation advice based on the number of past accurate and inaccurate advice. Then, it tries to adjust reputation advice according to its accuracy. However, TRAVOS assumes that sellers act consistently which may not be the case and further this model needs to go over an advisors past advice at each time when accuracy of the advice is to be estimated. For large number of advisors', managing vast amount of past advice is difficult.

An incentive mechanism to reinforce truthful reports [9] proposes a wage based scheme to encourage trustworthiness among peers. It gives high incentive for satisfactory reports and low incentive for non satisfactory reports. But, this scheme does not verify the truthfulness of received information.

A personalized approach for tackling unfair ratings [10] is suited for centralized reputation systems. To evaluate the advisors' trustworthiness, it uses a public and private reputation approach. It estimates advisors' credibility by using a probabilistic approach, and computes their trustworthiness based on their provided ratings.

Designing adaptive buying agents [12] approach is based on interacting with reputed advisors and based on the perceived error, it tries to adjust the reputation given by the witnesses. But, this model is based on allocating negative reputation also, which is disadvantageous in case of lowly rated participants exiting and re-entering the e-market with a fresh identity.

Evaluating rater credibility or reputation [21] uses a majority rating scheme to dilute the effect of unfair ratings by using standard deviation of all reported ratings and by adjusting the credibility of a service rater based on its past behaviour. This scheme suffers from the same problem like other systems from literature [1, 10, 12, 19] that are based on the assumption that mean, variance or standard deviation of the reported opinions would lead to an accurate assessment of the target agents' reputation. But in case the number of reported opinions is not very large, it may happen that the actual reputation lies near to the either end of the range of reported opinions rather than their mean. Moreover, predicting without the solid dependence on its own experience leads to a weak inference as far as accuracy of reputation is concerned.

An evidential model [2] for reputation management to select the fair advice considers three types of deceptions: complementary, exaggerative positive and exaggerative negative opinions. It uses an exaggeration coefficient to differentiate between exaggerative and complementary deceptive agents. This model uses the deviation of the ratings given by advisors to differentiate between fair and unfair advice.

Multi-Layer Cognitive Filtering [22] approach aggregates several parameters in computing the trustworthiness of advisors. It uses a two-layered filtering algorithm which first filters out neighbours with deceptive rating patterns and in the second layer it uses behavioural characteristics to cognitively derive actual intentions of the surviving agents of the previous layer.

The next section describes the proposed model for sharing reputation of the target seller agent among buyer agents. This model is relatively dynamic as it computes the weight of an advisor based on the dynamically changing parameters of e-market, like the number and percentage of honest past opinions and, the value of transaction. The amount of change in advisors' reputation is based on the value of transaction to emphasize that honest advice in a large value transaction is more important as compared to a similar advice in a relatively small value transaction.

## 3. Trust Enhancing Dynamic Reputation Sharing Model

The goal of a reputation system in an agent oriented e-commerce is to develop trustworthiness or the degree to which one agent/party has confidence in another within the context of a given purpose or decision [11]. A reputation system must ensure that after a number of transactions, the market reaches an equilibrium state and dishonest agents are weeded out of the market. To discourage dishonest agents in the e-market, an efficient reputation sharing model must filter out dishonest advice and, at the same time must also provide some incentive to honest advisors to encourage sharing of trustworthy i.e. honest opinions about reputation of others in an environment where all the participants are self interested.

## 3.1 Overview

The proposed strategy for sharing reputation of a seller among buyer agents is based on the reputation system in an e-market model [15, 16, 17, 18] having a set of buyers and sellers. In this model, sellers are divided into four categories, namely, reputed, non-reputed, dis-reputed and new sellers [16]. Buyers allocate reputation rating to sellers in the range [0,1]. At any given time t, a buyer preferably selects a seller from the list of reputed sellers but in no case, it selects a dis-reputed seller [16]. Before purchasing a good, the buyer computes expected value of the good being offered by each seller by using a dynamic seller selection strategy [17] and places an order to a seller that is offering the good with the highest expected value. After purchasing, once the buyer receives a good, it computes the actual value of that good. If actual value of the good is more than its expected value, the buyer increases the seller's individual reputation component; otherwise it decreases sellers' individual reputation. Further, the source buyer agent also solicits opinion from other reputed buyers' acting as advisors about the target seller agent and after filtering unfair advice it aggregates their opinions that form the shared reputation component. Finally, at time t+1, individual and shared reputations are combined to update overall reputation of the seller [16]. After the purchase of a good, individual reputation updated at time t+1 is based on overall reputation at time t to impress upon the fact that based on reinforcement learning, the updated overall reputation of a seller by a buyer at the end of previous transaction becomes the individual experience of that buyer agent in the next transaction. In addition to a buyer allocating a reputation to a seller, after each transaction, a seller agent also allocates reputation rating to the buyer with the purpose to increase buyer-seller satisfaction [15]. This reputation model is relatively dynamic in nature as it is sensitive to the changing parameters of the e-market like the value of a transaction and the changing experience of buyer-seller pair in the e-market [15, 16, 17, 18].

As depicted in Fig. 1, source buyer $b$ requests for opinion from other buyer agents $b_i$ where $b_i \neq b$ to obtain the reputation of the target seller agent $s$. After receiving the advice, it filters out dishonest advice and computes the weight of each opinion based on dynamically changing parameters like percentage and number of previous honest opinions from a particular buyer in the role of an advisor. It finally computes the aggregated shared opinion and updates the reputation of the advisor based on whether the opinion was evaluated to be honest or not.

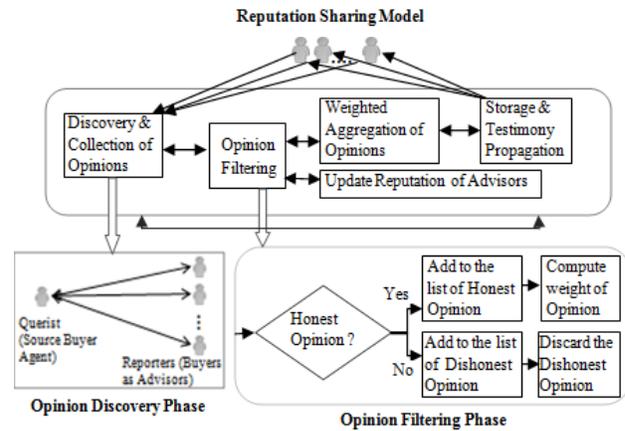

Fig. 1 Dynamic Reputation Sharing Model

There are certain issues in the process of collecting and aggregating opinion from advisors: buyers acting as advisors may provide unfair reputation of the target seller agent, buyers may not always be willing to share their opinion about the seller agent, so there should be some incentive for the honest advisors who are always ready to share their opinion with other buyer agents. Hence, before proceeding with an elaboration of how a group of advisors are selected, a possible desiderata that is important for an advisor is listed below.

i.  Transactional Experience: The opinion of those advisors would be more useful who have in-depth experience of the behaviour of the target seller agent in lieu of having participated in large number of transactions with the seller in the past.

ii. Behaviour Familiarity: It is beneficial to take the service of a reputed advisor repeatedly as it leads to increase in confidence and reliability due to its promising behaviour of successively providing honest opinions in the past.

iii. Reputation Computation Similarity: Advisors must compute the reputation of the target seller agent using the methodology similar to the one being adopted by the source buyer agent. This ensures that in the proposed dynamic reputation framework [8], after a transaction, all buyers compute the reputation of a seller using the same methodology [7, 8].

iv. Ideally, advisors should be as close as possible to the source buyer agent in the network topology i.e. at a maximum distance of not more than few hops. But this requirement is in contradiction to the availability of advice from sufficient number of reputed advisors i.e. advisors who have a good experience of the target seller agent and have also shown largely honest behaviour in previous transactions.

## 3.2 A Dynamic Reputation Sharing Methodology

As agents in an e-market are self-interested, they interact with each other to maximize their own goals. To ensure honest interaction amongst agents, a reputation system helps agents to estimate the trustworthiness of each other and choose the most reputable ones to interact with. However, a reputation system may be deceived by malicious agents that provide unfair advice for their personal gains. Hence, the problem of unfair ratings is fundamental and exists in almost every reputation system.

The proposed methodology to take care of the problem of unfair ratings is composed of two components: filtering unfair advice and, computing the reputation of advisors that reflects the analysis of their past behaviour in pursuing the goal of staying away from dishonest advisors. One of the essential requirements of such a system is the presence of willing advisors to participate in the opinion sharing process. This requires some sort of incentive for advisors as an encouragement to share their advice with other buyers. The proposed scheme in this paper provides incentive to honest advisors from two perspectives: first by associating a reputation with buyers exclusively for their contribution in the reputation sharing process as advisors. Second, with each successive honest/satisfactory opinion, there is an increase in the weight of an advisor as a reward for successive fair opinions in an environment where honesty is at premium.

The familiar folk theorem tells that cooperative behaviour can be sustained with infinite repetition and sufficient patient players [6]. This is possible in small societies where repeated and non-anonymous interaction is quite frequent, so that socially beneficial behaviour can develop. In this paper, the seeking of advice is restricted from a group of reputed advisors, which is a small group where entities have some knowledge of each other by virtue of their behaviour in previous transactions. This encourages cooperative and honest behaviour as compared to a large group of advisors with unknown identities.

The advice from other buyers is filtered by using the concept of "moment about an arbitrary point" [14] that compares the set of received opinions with the actual individual experience of the source buyer agent. The "second moment" is widely used and measures the "width" (in a particular sense) of a set of points in one dimension, or in higher dimensions measures the shape of a cloud of points as it could be fit by an ellipsoid [8]. Instead of finding the mean, variance or standard deviation from within the set of opinion data, this paper computes the distance of the individual reputation from each shared opinion about the target sellers' reputation by finding the second moment about a point representing individual experience of the source buyer with the target seller agent.

The second moment behaves just like variance but with respect to the individual experience of the source buyer agent. This is to emphasize the fact that reputation based on personal experience is a far more accurate representation of the accurate reputation value, from which the distance of the received opinions represented by the second moment is compared, in order to compute the maximum permissible error deviation that is acceptable for an opinion to be classified as honest. The set of opinions whose difference from the individual reputation is less than or equal to the value of second moment falls in the cluster of honest advice, and the opinions with difference more than the second moment falls in the cluster of dishonest advice.

In this model, B represents the set of buyers; A represents the set of buyers in the role of advisors where A∈B; S represents the set of sellers and, G the set of goods. Let $ar_t^b(b_i) \in [0,1)$ represents reputation rating of advisor $b_i$ at time t where i = 1,2,...,n; $sr_{t+1}^{b_i}(s)$ represents reputation of seller $s$ sent as opinion by advisor $b_i$ to the source buyer $b$; and $or_t^{others} \in [0,1)$ represents aggregated shared reputation (SR) i.e. the aggregated value of the opinion of other buyers about target seller $s$. At time t+1, buyer $b$ stores/remembers the reputation of all the advisors, with whom it interacted at time t in the past. Each buyer maintains four categories of advisors as defined below.

(i) $A_R^b$ : Advisors in the reputed list of buyer $b$, i.e. $ar^b(b_i) \geq \hat{\theta}_{b_i}^b$, where advisor $b_i \in A_R^b$, $\hat{\theta}_{b_i}^b$ is the reputation threshold of advisor $b_i$ for i=1,2,...,n for source buyer agent $b$ and $0 < \hat{\theta}_{b_i}^b < 1$.

(ii) $A_{NR}^b$ : Advisors in the non-reputed list of buyer $b$, i.e. $\check{\theta}_{b_i}^b < ar^b(b_i) < \hat{\theta}_{b_i}^b$ where advisor $b_i \in A_{NR}^b$.

(iii) $A_{DR}^b$ : Advisors in the dis-reputed list of buyer $b$, i.e. $0 < ar^b(b_i) \leq \check{\theta}_{b_i}^b$ where $b_i \in A_{DR}^b$, $\check{\theta}_{b_i}^b$ is the dis-reputation threshold and $0 < \check{\theta}_{b_i}^b < \hat{\theta}_{b_i}^b$.

(iv) $A_{NewR}^b$ : Advisors those are new to buyer $b$ in the market, initially $ar^b(b_i) = 0$. A new advisor $b_i$ remains in this list until its reputation crosses the dis-reputation threshold $\check{\theta}_{b_i}^b$.

Let $a$ be the minimum threshold on the number of advisors whose opinion is necessary before computing a consolidated shared reputation rating of the target seller agent. This is done to add to the fairness of shared opinion component. Further, let $b$ be the source buyer agent that requests for opinion about the past performance of the target seller agent $s$ from other buyers at time t+1 and, in return, a subset of advisors respond. The detailed methodology is divided into two parts and is elaborated ahead.

Part I:

1. The source buyer agent $b$ solicits opinion about the target seller agent s from the set of buyer agents who belong to its list of reputed advisors i.e. buyers who provided satisfactory opinions in the past interactions to buyer $b$. If the number of advisors sharing their opinion about the reputation of the target seller agent $s$ falls below the minimum threshold for the number of advisors then the buyer $b$ seeks opinion from the set of non-reputed advisors but in no case the buyer would choose a dis-reputed advisor. In addition, with a small probability ð, buyer $b$ would solicit opinion from previously unknown advisors who would join the set of new advisors i.e. $A_{NewR}^{b}$. Initially the value of ð is 1 and it decreases over time to some minimum value defined by buyer $b$.

2. Accept the advice from a subset of advisors $b_i \in A$ for i =1,2,...,n, who respond with their opinion $sr_{t+1}^{b_i}(s)$, that represents reputation of seller $s$ sent as opinion by advisor $b_i$ to the buyer $b$. Ignore, if there is an opinion from any of the dis-reputed advisors.

Part II:

3. To filter out unfair opinions, find the second moment about the individual reputation value of the target seller agent to compute the maximum permissible deviation from the individual reputation rating. This is to ensure that opinions too diverging from its own experience about the target seller agent are discarded and not taken into consideration while computing the overall shared reputation component $or_{t+1}^{others}(s)$. Based on shared reputation $sr_{t+1}^{b_i}(s)$ and individual reputation of seller $s$ i.e. $r_{t+1}^{b}(s)$, the equation to compute second moment $m_2$ about an arbitrary point denoted by the individual reputation component $r_{t+1}^{b}(s)$ is illustrated in Eq. (1).

$$m_2 = \frac{\sum_{i=1}^{n}(sr_{t+1}^{b_i}(s) - r_{t+1}^{b}(s))^2}{n} \quad (1)$$

4. All the opinions for which the difference $sr_{t+1}^{b_i}(s) - r_{t+1}^{b}(s) \leq m_2$, denote the opinion as fair/honest and the set of advisors as honest advisors represented by $\tilde{H}$. All the opinions for which the difference $sr_{t+1}^{b_i}(s) - r_{t+1}^{b}(s) > m_2$, denote these opinions as unfair and the set of advisors as dis-honest advisors $\tilde{D}$. Now, filter out the opinions of dis-honest advisors $\tilde{D}$ from the set of advice to be considered for aggregation.

5. Find the weightage $w_{b_i}$ of each advisor $b_i$ for i = 1, 2,..., n based on its past behaviour by using Eq. (2).

$$w_{b_i} = 1 - e^{-\kappa \cdot \zeta} \quad (2)$$

In Eq. (2), $\kappa$ is constant in the range (0,1], and $\zeta$ represents the consolidated value of various parameters of advisors' past behaviour as shown in Eq. (3).

$$\zeta = \acute{a} * \%OfHPO + \acute{\varepsilon} * NoOfHPO * \mathcal{I} \quad (3)$$

Where $\%OfHPO$ represents the percentage of Honest Past Opinions from the advisor and $\acute{a}$ is a constant in the range (0,1] that represents the effect of $\%OfHPO$ on advisors weightage. Further, $NoOfHPO$ represents the number of Honest Past Opinions from the advisor and $\acute{\varepsilon}$ is a constant in the range (0,1] that represents its effect on advisors' weightage. Also, $\mathcal{I}$ represents the incentive of being a fair advisor with the initial value of $\mathcal{I} = 1$. With each successive honest advice from a particular advisor, the value of $\mathcal{I}$ is incremented by a small factor say 0.01 to give greater incentive to a faithful advisor. The product of $NoOfHPO$ and $\mathcal{I}$ ensure that the effect of incentive provided to advisor for being honest changes dynamically and increases by a greater margin with each successive honest advice. The equation for %ofHPO is given below in Eq. (4).

$$\%ofHPO = \frac{No.ofHonstOpinions}{TotalNo.ofOpinons} * 100 \quad (4)$$

6. Find the aggregated shared reputation of the seller $s$ i.e. $or_{t+1}^{others}(s)$ by using Eq. (5) below.

$$or_{t+1}^{others}(s) = \frac{\sum_{i=1}^{n} w_{b_i} * sr_{t+1}^{b_i}(s)}{\sum_{i=1}^{n} w_{b_i}} \quad (5)$$

7. Update the reputation $ar_{t+1}^{b}(b_i)$ of each advisor $b_i$, whose opinion was classified as honest i.e. $b_i \in \tilde{H}$ using reinforcement learning as shown in Eq. (6).

$$ar_{t+1}^{b}(b_i) = ar_t^{b}(b_i) + \Omega(1 - ar_t^{b}(b_i)) \quad (6)$$

$$\text{Further,} \quad \Omega = 1 - e^{-\Lambda \cdot x} \quad (7)$$

Where $\Omega$ is the advisors' reputation increase factor and $x$ represents the value of transaction for which the opinion about the target seller agent is sought from the advisor by the source buyer agent. This makes the amount of increase in the advisors' reputation to be truly dynamic as it is monotonically proportional to the value of transaction $x$. Hence, the amount of advisors' reputation enhanced for a large value transaction would be greater than its enhancement in case of a relatively small value transaction.

8. Update the reputation of each dishonest advisor $b_i \in \tilde{D}$ as shown below in Eq. (8).

$$ar_{t+1}^{b}(b_i) = ar_t^{b}(b_i) - \varphi(1 - ar_t^{b}(b_i)) \quad (8)$$

Further, $\varphi = p * (1 - e^{-\Lambda \cdot x})$ (9)

Where $\varphi$ represents the advisors' reputation decrease factor, $\Lambda$ is a constant in the range between 0 to 1, and $p$ represents the penalty factor for decrease in reputation and $p \geq 1$ to emphasize that the penalty of being dishonest must always be at least equal to or greater than the corresponding reward for being honest. This is based on the convention that reputation is hard to build but easy to tear down. The actual value of $p$ can be decided by the domain experts. The process of decreasing the dishonest advisors' reputation is also dynamic as it is monotonically proportional to the value of a transaction based on the convention that dishonest behaviour in a relatively large value transaction is more devastating and hence must be discouraged to a greater extent than similar behaviour in a small value transaction.

9. Finally, based on step 7 or step 8, for i=1,2,..,n, update the sets of reputed, non-reputed, dis-reputed and new advisors i.e. $A_R^b$, $A_{NR}^b$, $A_{DR}^b$ and $A_{NewR}^b$ as:

If $b_i$ is not a reputed advisor, and $ar_{t+1}^b(b_i) \geq \hat{\theta}_{b_i}^b$,

$$A_R^b = A_R^b \cup \{b_i\}.$$ (10)

If $b_i$ is a reputed advisor, and $ar_{t+1}^b(b_i) < \hat{\theta}_{b_i}^b$,

$$A_R^b = A_R^b - \{b_i\}.$$ (11)

If $b_i$ is not a dis-reputed advisor, and $ar_{t+1}^b(b_i) \leq \hat{\theta}_{b_i}^b$,

$$A_{DR}^b = A_{DR}^b \cup \{b_i\}.$$ (12)

If $b_i$ is not non-reputed, and $\theta^b < ar_{t+1}^b(b_i) < \hat{\theta}_{b_i}^b$,

$$A_{NR}^b = A_{NR}^b \cup \{b_i\}.$$ (13)

Finally, if $b_i$ is a new advisor, and, if $ar_{t+1}^b(b_i) > \hat{\theta}_{b_i}^b$,

$$A_{NewR}^b = A_{NewR}^b - \{b_i\}.$$ (14)

As all participants in the reputation system are self interested, hence there must be some incentive for a buyer to participate in the opinion sharing process. This work provides incentives to an honest advisor by increasing its reputation i.e. trustworthiness and also by dynamically increasing the weight of its opinion with each successive honest advice. It is based on the convention that wrong advice is worse than no advice, as its purpose is to subvert the main goal of reputation systems by artificially manipulating the reputation of the target seller agent through collusion, to mislead the source buyer agent that is seeking opinion. Hence, this methodology provides for penalising the dishonest advisors by reciprocatively decrementing their reputation as advisors.

Further, to ensure that there is always an availability of honest advisors, this model is based on the premise that a buyer must always respond to a request for an advice from any other buyer who belongs to its reputed advisors' list. This is based on the convention, that good behaviour in case of honest non competitive peers who engage in repeated transactions is a win-win game. This is mutually beneficial, as if buyer x always responds to a request from y, and in return y also responds to x's request for advice, then the need of seeking opinion from relatively unknown and potentially dishonest advisors is minimised.

To summarize, the main functions of the proposed dynamic reputation sharing methodology with an intention to enhance trustworthiness of the system by filtering out dishonest opinions and by allocating reputation to advisors, are illustrated using a flowchart in Fig. 2.

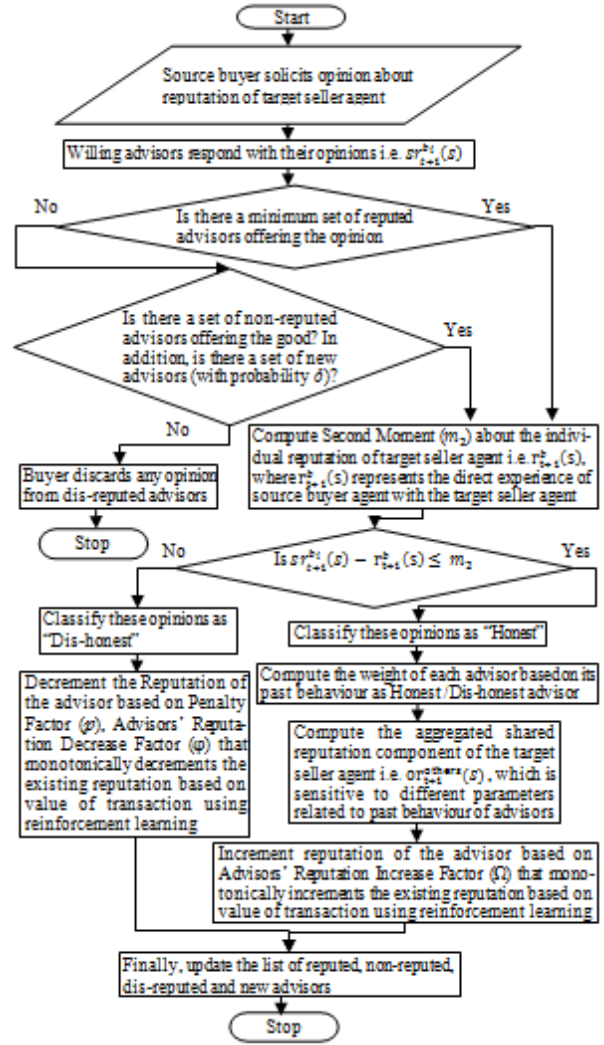

Fig. 2  A Dynamic Reputation Sharing Methodology

The proposed model is sensitive to the dynamically changing parameters of e-market, like the value of transaction and the past behaviour of advisors, with the purpose to encourage advisors who contribute to the trustworthiness of the reputation system and e-market environment by repeatedly providing honest opinions.

## 4. Case Study

To illustrate the application of the proposed model for sharing reputation of sellers among buyers in order to enable them to make an informed decision about selecting honest advice, a case study was conducted by simulating an electronic marketplace with eight participants as buyers and five as sellers, i.e. B = {$b_i$ where $i$ = 1...8}, set A∈B, and S = {$s_j$ where $j$ = 1...5}, where B is the set of buyers, A is the subset of buyers playing the role of advisors, and S is the set of sellers in the marketplace for good $g$. The proposed model was simulated in MATLAB and is illustrated ahead.

A situation was investigated where buyer $b_2$ requested for an opinion about the reputation of a seller $s_4$ from other buyers, after purchasing the good $g$. The buyers $b_1$, $b_3$, $b_4$, $b_5$ and $b_7$, acting as advisors were known to buyer $b_2$, although only four buyers were in its overall reputed list of advisors i.e. $A_R^{b_2}$ = {$b_j$ where $j$ = 1,3,5,7}. At time t+1 the source buyer agent $b_2$ stored the number of honest past opinions and the percentage of honest past opinions as shown in Table 1 below.

Table 1: Reputation of Buyers as Advisors

| $b_i$ | $b_1$ | $b_3$ | $b_4$ | $b_5$ | $b_7$ |
|---|---|---|---|---|---|
| %OfHPO | 76.92 | 82.3 | 46.8 | 79.4 | 90 |
| NoOfHPO | 20 | 14 | 8 | 11 | 9 |

Further, $\hat{\theta}_{b_i}^{b_2}$ = 0.38, $\tilde{\theta}_{b_i}^{b_2}$ = 0.15, e = 1.01, $\acute{a}$ = 0.6, $\acute{\varepsilon}$ = 0.3, $\kappa$ = 0.95, $\underline{a}$ = 2, $\Lambda$ = 0.001 and, penalty factor $\underline{p}$ = 1.5. Based on the source buyer agent $b_2$'s request, the set of opinions received from other buyers who responded with their advice about the reputation $sr_{t+1}^{b_i}(s_4)$ of seller $s_4$ and, the current reputation of different advisors $ar_t^{b_i}(b_2)$ being stored by the buyer $b_2$ that is based on their past behaviour are depicted in Fig. 3 ahead.

The opinion from advisor $b_4$ is ignored by the source buyer agent $b_2$ as $b_4$'s reputation as advisor i.e. ar_b4 = 0.31 is less than reputation threshold of $\hat{\theta}_{b_i}^{b_2}$= 0.38. The reputation component $r_{t+1}^{b_2}(s_4)$ reprsenting the target seller $s_4$'s reputation based on source buyer agent $b_2$'s individual experience was 0.389.

Fig. 3 Shared opinion of seller $s_4$'s reputation from different advisors, and current reputation of advisors in $b_2$'s Memory

Using Eq. (1), second moment $m_2$ representing distance between the opinions by the set of reputed advisors and the individual experience of buyer $b_2$ about the reputation of seller $s_4$ was computed to be 0.0092. Also, using step 4, part II of the methodology, opinions for which the difference $sr_{t+1}^{b_i}(s_4) - r_{t+1}^{b_2}(s_4) \leq m_2$ were classified to be "Honest" and others as "Dishonest" using MATLAB as shown in Fig. 4.

Fig. 4 Filtered opinion of Advisors' as Honest or Dishonest

The value of $\mathcal{I}$ representing incentive for honest opinions for a $\mathcal{I}$ incremental value of 0.01 after each transaction is shown below in Table 2.

Table 2: Value of $\mathcal{I}$ as incentive for honest opinions

| $b_i$ | $b_1$ | $b_3$ | $b_7$ |
|---|---|---|---|
| $\mathcal{I}$ | 1.2 | 1.14 | 1.09 |

Using Eq. (3), representing the consolidated value of the past behaviour of the advisors ($\zeta$), the weight of each reputed advisors' opinion ($w_{b_i}$) was computed, and was found to be monotonically proportional to the past behaviour of various advisors i.e. the advisors with successively honest past behaviour were allocated higher weight as compared to others as shown in Fig. 5 below

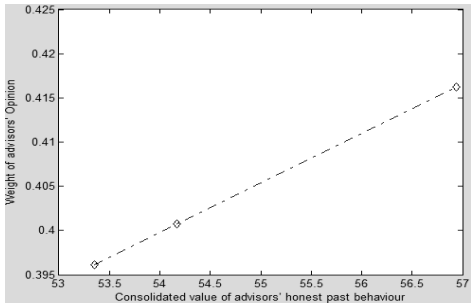

Fig. 5 Weight of advisors' opinion varies monotonically w.r.t. consolidated value of their past behaviour ($\zeta$)

Now, based on the newly computed weight of each advisor ($w_{b_i}$), the aggregated shared reputation of the seller $s_4$ i.e. $or_t^{others}(s)$ was computed and represented as ar_others = 0.3875 in MATLAB as shown in Fig. 6 below. Further, based on Eq. (7), the value of $\Omega$ was computed as:

$$\Omega = 1 - 1.01^{(-0.001*1800)} = 0.01775. \quad (15)$$

And, using Eq. (9), the value of $\varphi$ was computed as:

$$\varphi = 1.5 * \left(1 - (1.01)^{(-0.001*1800)}\right) = 0.02663. \quad (16)$$

Using Eq. (6) and Eq. (8), updated reputation of different advisors i.e. $ar_{t+1}^{b_i}(b_2)$ was computed using MATLAB. The aggregated shared reputation 'or_others' of seller $s_4$, weights of honest advisors and the updated reputation of different advisors is illustrated in Fig. 6.

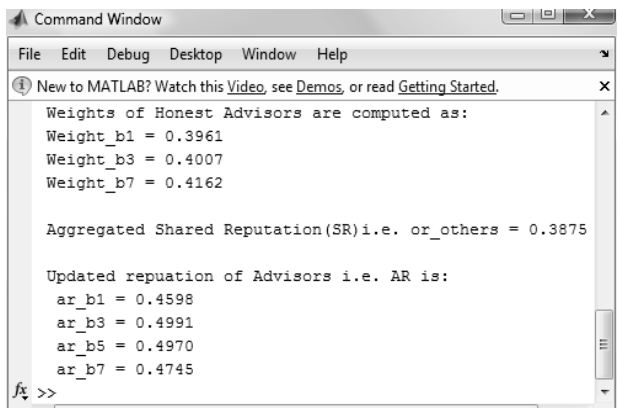

Fig. 6 Aggregated Shared Reputation of seller $s_4$ and Updated Reputation of Advisors

Finally, the aggregated shared reputation component i.e. or_others equalling 0.3875 was communicated for its use in updating the overall reputation of seller $s_4$ for reference in any future transactions by buyer $b_2$ with seller $s_4$.

## 5. Conclusions

This paper proposed a dynamic strategy for buyer agents to enable a buyer to make an informed decision of selecting honest advisors by filtering unfair opinions in order to generate an aggregated shared reputation component. The strategy provided an incentive to honest advisors by associating the concept of reputation with advisors using reinforcement learning with the purpose to improve the trustworthiness among participants in the e-market environment.. The proposed model is relatively dynamic as it computes the weight of an advisors' opinion based on the changing parameters of e-market, like the number and percentage of honest past opinions. Further, the amount of change in advisors' reputation is based on the value of a transaction to emphasize that honest advice in a large value transaction is more important as compared to a similar advice in a relatively small value transaction.

**Vibha Gaur:** She is PhD from the department of Computer Science, Delhi University. She is working as Associate Professor in Delhi University and has a teaching experience of about 14 years. She has authored more than 20 papers in various international conferences and journals. Her current research interests include artificial intelligence, information systems, trust, e-commerce, software quality and requirement engineering.

**Neeraj Kumar Sharma:** PhD student at Delhi University. He is also working as Assistant Professor in Delhi University and has a teaching experience of about 8 years. He has published two booklets pertaining to MCA syllabus of IGNOU in the subjects Artificial Intelligence and Algorithms in 2005 and 2006. He has presented two papers in international conferences by ACEEE (ACS 2010) and Springer (ACC 2011). He has also published two papers in international journals, one paper in International Journal on Recent Trends in Engineering & Technology and one in International Journal of Computer Science Issues (IJCSI, Vol.8, Issue4, July2011).

**Punam Bedi:** She is PhD from the department of Computer Science, Delhi University. She is working as Associate Professor in the Department of Computer Science at Delhi University and has a teaching experience of more than 25 years. She has authored more than 50 papers in various international conferences and journals. Her current research interests include artificial intelligence, information systems, trust, recommender systems, human-machine interaction and different branches of software engineering.